\documentclass[psf,epsf,twocolumn,floatfix]{revtex4}
\usepackage{graphics}
\usepackage{graphicx}
\usepackage{dcolumn} 
\usepackage{bm}
\usepackage{epsfig}
\newcommand{\hem}{Fe$_2$O$_3$}
\newcommand{\ahem}{$\alpha$-Fe$_2$O$_3$}
\newcommand{\ilm}{FeTiO$_3$}

\newcommand{\fetw}{Fe$^{2+}$}
\newcommand{\feth}{Fe$^{3+}$}
\newcommand{\tith}{Ti$^{3+}$}
\newcommand{\tif}{Ti$^{4+}$}
\newcommand{\otw}{O$^{2-}$}
\newcommand{\eg}{{\sl e.g.}}
\newcommand{\etal}{{\sl et al.}}

\newcommand{\lmh}{{\sl lamellar magnetism hypothesis}}
\begin{document}
\title{Interface magnetism in Fe$_2$O$_3$/FeTiO$_3$-heterostructures }
\author{Rossitza Pentcheva} \email{pentcheva@lrz.uni-muenchen.de}
\author{Hasan Sadat Nabi} \affiliation{Department of Earth and
  Environmental Sciences, University of Munich, Theresienstr. 41,
  80333 Munich, Germany} \date{\today}

\begin{abstract}
  To resolve the microscopic origin of magnetism in the Fe$_2$O$_3$/FeTiO$_3$-system, we have performed
density functional theory calculations taking into account on-site Coulomb repulsion. By
  varying systematically the concentration, distribution and charge state of Ti in a
  hematite host, we compile a phase diagram of the
  stability with respect to the end members and find a clear preference to form layered
  arrangements as opposed to solid solutions. The charge mismatch at the interface is accommodated through
  Ti$^{4+}$ and a disproportionation in the Fe contact layer into Fe$^{2+}$, Fe$^{3+}$, leading to uncompensated
moments in the contact layer and giving first theoretical evidence for the {\sl lamellar magnetism
  hypothesis}. This interface magnetism is associated with impurity levels
in the band gap showing halfmetallic behavior and making Fe$_2$O$_3$/FeTiO$_3$ heterostructures
prospective materials for spintronics applications.
\end{abstract}
\maketitle

A challenge of today's materials science is to design ferromagnetic
semiconductors operating at room-temperature (RT) for spintronics devices.
Most of the efforts concentrate on homogeneous doping of semiconductors
with magnetic impurities~\cite{Coey05,MacDonald05,Chambers06a,Kuroda07}, but the
interfaces in complex oxides prove to be another source of novel
behavior~\cite{ohtomo2002,brinkman07,reyren07}.
The unique magnetic properties of the hematite-ilmenite system~\cite{ishikawa57,ishikawa58,carmichael61}
(a canted antiferromagnet and a RT paramagnet, respectively)
currently receive revived interest as a possible cause of
magnetic anomalies in the Earth's deep crust and on other planets~\cite{McEnroe04} as well as
for future device applications~\cite{takada07,hojo06,Chambers06a}.

\begin{figure}[b]
\begin{center}
  \rotatebox{0.}{ \includegraphics[scale=0.40]{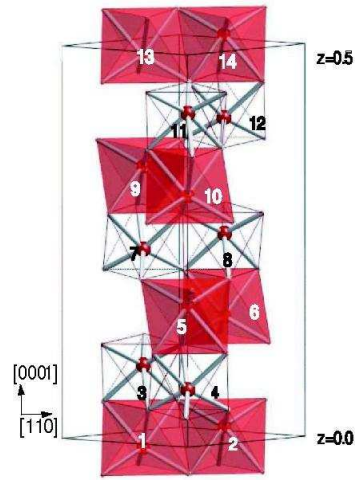}}
\end{center}
\caption{\label{fig:struct}  (Color online) Crystal structure  of \ilm, showing half of the
  60 atom unit cell. The cation sites are numbered,
 oxygen occupies the edges of the octahedra.  }
\end{figure}

Both hematite ($a=5.035$\AA, $c=13.751$\AA~\cite{attfield91}) and ilmenite
($a=5.177$\AA, $c=14.265$\AA~\cite{harrison00})
crystallize in a corundum(-derived) structure
shown in Fig.~\ref{fig:struct} where the oxygen ions form a distorted hexagonal
close packed lattice and the cations occupy $2/3$ of the octahedral
sites. In \ahem~(space group $R\bar{3}c$) there is a natural modulation of electronic density
along the [0001]-direction where negatively charged 3O$^{2-}$ layers
alternate with positively charged 2\feth~layers.  At RT the magnetic moments of
subsequent iron layers couple antiferromagnetically (AFM) in-plane with a small
spin-canting, attributed to spin-orbit
coupling~\cite{dzialoshinski,moriya,sandratski}.  In
ilmenite, \ilm, Fe- and Ti-layers alternate, reducing the symmetry to $R\bar{3}$
and the corresponding sequence
is 3\otw/2\fetw/3\otw/2\tif~with AFM coupling between the Fe layers
and $T_{\rm N}=$56-59~K~\cite{Mcdon91}.

At the interfaces (IFs) in hematite-ilmenite exsolutions, charge
neutrality is disrupted. One way to balance the
excess charge at the interface is by a
disproportionation in the Fe layer, now
becoming mixed \fetw~and \feth. This {\sl lamellar magnetism hypothesis (LMH)}
was proposed by Robinson
\etal~\cite{nature04} based on bond valence models
and kinetic Monte Carlo simulations (kMC) with empirical chemical and
magnetic interaction parameters.  The increased technological interest
in this system calls for an atomistic material specific understanding that can only be obtained
from first principles calculations. A previous density functional
theory (DFT) study within the generalized gradient approximation
(GGA) found no evidence for the LMH~\cite{burton05}. However, electronic correlations, not included in
the local (spin-) density approximation (LSDA) or GGA of DFT, play an important role in
transition metal oxides.  Such effects were considered recently within
LSDA+U~\cite{butler05} or using hybrid functionals~\cite{droubay07}
for single Ti impurities in hematite, however layered
arrangements and interfaces were not addressed.

In this paper we have performed DFT calculations including a Hubbard U~\cite{anisimov93}
for the end members \hem~and \ilm, as well as their interfaces and
solid solutions (SS). By varying systematically the concentration,
distribution and charge state of Ti incorporated in a \ahem-host,
we explore different scenarios for the charge compensation mechanism
 and its consequences for the magnetic and electronic
behavior. Finally, we compile a phase diagram of
the stability of the different configurations with respect to the end
members as a function of Ti-doping also taking into account the effect
of strain.

Our DFT-GGA~\cite{pbe96} calculations are performed using the
all-electron full-potential augmented
plane waves (FP-APW) method as implemented in  \textsf{WIEN2k}~\cite{wien}.
Electronic correlations are considered within the fully localized limit
(LDA+U)~\cite{anisimov93}. The systems are modeled in the hexagonal primitive unit cell (shown in
Fig.~\ref{fig:struct}) containing 30 and 60 atoms.  For these 24 and 15 $k$-points in the irreducible part of the Brillouin zone were used, respectively. Inside the muffin tins ($R_{\rm MT}^{\rm Fe,Ti}=1.80\,
  \textrm{bohr}$, $R_{\rm MT}^{\rm O}=1.60\,\textrm{bohr}$) wave functions are expanded in spherical
  harmonics up to $l_{\rm max}^{\rm wf}=10$ and non-spherical
  contributions to the electron density and potential up to $l_{\rm
    max}^{\rm pot.}=6$ are used. The energy cutoff for the plane wave
  representation in the interstitial is $E_{\rm max}^{\rm wf}=19$~Ry
  for the wave functions and $E_{\rm max}^{\rm pot.}=196$~Ry for the
  potential.  These convergence parameters ensure
a numerical accuracy of energy differences better than 0.01eV/60 atom cell.
A full structural optimization of internal parameters has been performed~\cite{ldauforce}.

As a starting point we have modeled the end
members \hem\ and \ilm. In agreement with previous
calculations~\cite{rollmann04,butler05}, GGA+U considerably improves
the band gap of hematite from  0.43 eV (GGA) to 2.2~eV for U=6~eV and J=1~eV
in close agreement with measured values of
2.14-2.36~eV~\cite{benjelloun84,chang72}. Also the type of band gap changes
from a Mott-Hubbard  between Fe$3d$- Fe$3d$ states to charge
transfer after the scheme of Zaanen \etal~\cite{Zaanen85} between
occupied O$2p$ and empty Fe$3d$-states.

For ilmenite GGA incorrectly predicts a metallic state~\cite{harrison05}, hence
the inclusion of Hubbard U ($U=8$~eV, $J=1$~eV) is decisive to obtain a Mott-Hubbard gap
of 2.18~eV ($\Delta^{exp}=2.54$~eV~\cite{ishikawa58}) between
the occupied Fe$d_{z^2}$-orbital and the
unoccupied Fe$3d$ states in one spin channel (all Fe$3d$ orbitals being
occupied in the other spin-channel). These $U$ and
$J$ values are used both on Fe and Ti in the following. A
\feth/\tith charge arrangement lies 0.63~eV/p.f.u. above the ground state
\fetw/\tif. For Fe$^{2+}$Ti$^{4+}$O$_3$ the AFM and FM coupling between
Fe-layers is nearly degenerate, consistent with the low T$_N$.

  \begin{table}
 \caption{\label{tab:enmm} Relative stability (eV/60 at. cell) of the different cation
arrangements in the \tif, \fetw, \feth charge state (strained at the \hem\ lattice parameters) with respect to
 the most stable configuration whose energy is set to 0.0 eV.
The positions occupied by Ti are denoted as subscripts
according to Fig.~\ref{fig:struct} (SSL: spin-sublattice).  The total magnetic
moment, electronic behavior (hm/m denotes halfmetallic/metallic]
as well as \tif-O and \fetw-O distances at the interface
are also displayed.
}
 \begin{ruledtabular}
  \begin{tabular}{lcccrr}
   System &$\Delta E$ &$d_{{\rm Ti}^{4+}{\rm -O}}$ & $d_{{\rm Fe}^{2+}-{\rm O}}$& M$_{\rm tot}$  & $\Delta$\\
      &(eV) & \AA & \AA &$(\mu_{\rm B})$& (eV)\\
 \hline
\multicolumn{4}{c}{Ilm$_{17}$}  \\
T$_{1,2}$: 1 Ti-layer & 0.0 & 1.90 & 2.08& -8.0& {\sl hm}\\
T$_{7}$: single impurity   &  0.20 & 1.93 & 2.02& 12.0& {\sl hm}\\
\multicolumn{4}{c}{Ilm$_{33}$}  \\
 T$_{1,2,13,14}$: 1 Ti-layer  & 0.36 & 1.90 & 2.08 & -16.0& {\sl hm}\\
 T$_{3,4}$: 2 Ti-layers & 0.00 & 1.89 & 2.08 & 16.0& {\sl hm}\\
 T$_{5,10}$:  same SSL & 0.19 & 1.86 & 2.04 & -16.0 & {\sl hm}\\
 T$_{5,7}$:  different SSL & 0.41 & & & 0.0 &{\sl m}\\
 T$_{5,8}$:  different SSL & 1.24 & 1.86 & 2.06 & 0.0& {\sl hm}\\
 \multicolumn{4}{c}{Ilm$_{50}$}  \\
  T$_{1,2,5,6}$: 3 Ti-layers & 0.00  & 1.90 & 2.07 & -24.0&  {\sl hm}\\
  T$_{3,4,7}$: Ti-Fe@IF & 1.23  & 1.96 & 2.00 & 8.0&  {\sl hm}\\
 \multicolumn{4}{c}{Ilm$_{66}$}  \\
  T$_{3,4,11,12}$: 4 Ti-layers & 0.00 & 1.90& 2.07 & 16.0& {\sl hm}\\
  T$_{3,6,9,12}$: SS & 0.95 & 1.92& 2.02 & 0.0& {\sl hm}\\
 \end{tabular}
\end{ruledtabular}
\end{table}

\begin{figure}[t]
  \rotatebox{0.}{ \includegraphics[scale=0.38]{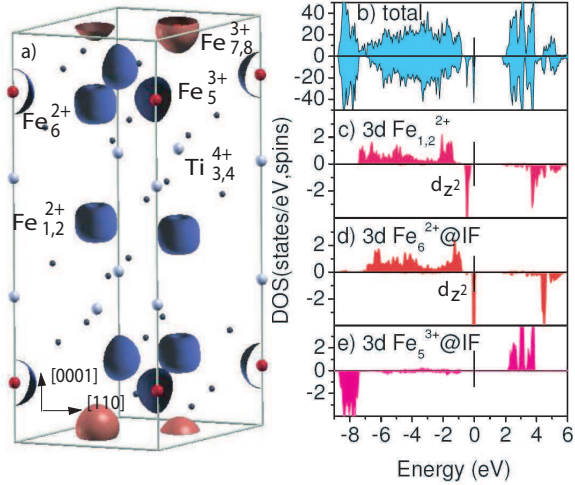}}
 \caption{\label{fig:cdndos_hemt3t4} (Color online) a) Spin density distribution and b-e) total and projected density of states of Ti double layer in hematite (T$_{3,4}$) with \fetw$_{1,2}$, \tif$_{3,4}$ and a disproportionated \feth$_{5}$, \fetw$_{6}$ layer at the interface. The position of the Fermi level is set to 0.0 eV and denoted by a short line. In a) positions of Ti, Fe and O are marked by light grey, red and small black circles, respectively.
}
\end{figure}

In the following we vary the concentration and distribution of Ti in a
\hem-host.  The positions of the Ti-ions are given as subscripts and
follow the notation of Fig.~\ref{fig:struct}, ~\eg~T$_{3,4,7,8,11,12}$
describes pure ilmenite.  Table~\ref{tab:enmm} contains the energetic
stability, structural, magnetic and electronic properties of different cation arrangements and concentrations.
We start the discussion with Ilm$_{33}$ which
corresponds to four Ti-ions out of 24 cations in the 60-atom cell.
We find that the formation of a compact
ilmenite-like block with a Fe-layer sandwiched between two
Ti-layers (T$_{3,4}$), is by 0.36~eV more favorable than incorporation of single Ti layers
in the hematite host (T$_{1,2,13,14}$).  The spin density of T$_{3,4}$
plotted in Fig.~\ref{fig:cdndos_hemt3t4}a) shows that the central Fe-layer turns into
\fetw~and the charge mismatch at the IF is compensated by \fetw,\feth\ in the contact
layer, giving theoretical evidence from
first principles for the {\sl lamellar magnetism} hypothesis of
Robinson \etal~\cite{nature04}. Our GGA+U calculations show that
\tif$_4$ shares faces with \feth$_5$, while \fetw$_6$ shares faces
with \feth$_8$ from the next hematite layer. Such a configuration was
proposed  using bond-valence sums~\cite{robinson06}
and kMC~\cite{harrison06} only after considering both chemical and magnetic interactions.

The formation of layered arrangements (T$_{3,4}$) is favored compared to a more random
distributions with 50\% substituted cation layers
(\eg~T$_{5,10}$, T$_{5,7}$ or T$_{5,8}$). With respect to magnetism, each Ti ion adds a magnetic moment of
4$\mu_B$ independent of whether the extra electron is localized at Ti
(\tith) or at a neighboring Fe (\fetw). In solid solutions the total magnetic
moment depends on the site and sublattice where Ti is built in.
We find that incorporation in the same spin-sublattice T$_{5,10}$ (which maximizes the magnetic moment)
 is favored by 0.22~eV compared to the AFM
T$_{5,7}$. Taking into account local lattice relaxations enhances the energy gain compared
to previous calculations by Velev \etal~(0.08~eV)\cite{butler05}.
Still, some degree of Ti disorder is likely in quickly cooled
samples, reducing the expected  magnetization as observed by
Chambers \etal~\cite{Chambers06a} (0.5$\pm 0.15\mu_B$/Ti for $x_{\rm Ti}= 0.15$). For higher Ti
concentrations and longer annealing steps the formation of the thermodynamically
more stable layered ferrimagnetic phase is expected, consistent with the strong correlation between
cation order and ferrimagnetism found in annealed samples~\cite{ishikawa57,takada07}
and the saturation magnetization of $3\mu_B$/mol measured in epitaxial films with
$x_{\rm Ti}\leq 0.63$~\cite{hojo06}. While below the ordering temperature of ilmenite only ilmenite
lamella with an odd number of Ti-layers are expected to carry a non-zero magnetization, above 56~K
the net magnetic moment will be solely due to the uncompensated magnetic moments in the contact
layers, independent of the number of Ti-layers within the paramagnetic ilmenite lamella.

Concerning the electronic properties, doping \hem~with Ti leads to impurity levels in
the band gap arising from the occupied $d_{z^2}$-orbital of \fetw~ions in the contact
layer. The density of states plotted in Fig.~\ref{fig:cdndos_hemt3t4}b-e) shows that these
states are pinned at the Fermi level, leading to fully spin-polarized carriers and
half-metallic behavior for  T$_{3,4}$. This trend is robust with respect to $U$~\cite{U=6} and is observed for most of the studied
cation concentrations and arrangements after structural relaxation.
Experimentally, semiconducting behavior and a drop in resistivity of several orders
w.r.t. the end members is measured~\cite{ishikawa58,takada07,hojo06,Chambers06a} with
values suggesting localized rather than itinerant carriers
consistent with the picture we obtain from LDA+U.
 \begin{figure}[t]
  \rotatebox{0.}{ \includegraphics[scale=0.72]{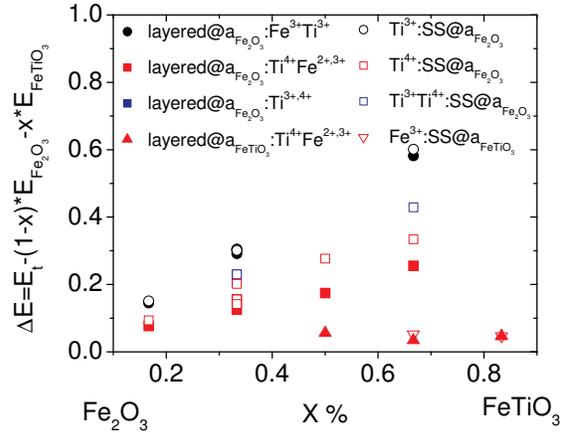}}
 \caption{\label{fig:Ex} (Color online)  Formation energy (eV/p.f.u) with respect to the end members as a function of ilmenite concentration $x$. Layered arrangements (solid solutions) are denoted by full (open) symbols. 
 }
 \end{figure}

Next we turn to the energetic stability as a function of the
Ti-concentration displayed in Table~\ref{tab:enmm} and the phase
diagram in Fig.~\ref{fig:Ex}. Charge compensation through \tif\ and disproportionation of iron into \fetw, \feth\ is strongly
favored compared to compensation involving \tith, especially after optimization of the internal
structural parameters.
Moreover, the formation of layered configurations (full symbols) is preferred
over disordered arrangements
(empty symbols) except for
very high ($>83$\%) and very low concentrations ($<17$\%), consistent the miscibility gap
from thermodynamic data (\eg~\cite{burton05,McEnroe02b}).
The linear increase of formation energy in the range between
Ilm$_{17}$ and Ilm$_{66}$ indicates that straining Ti doped \hem~to the \hem~lattice
parameters gets increasingly unfavorable with growing $x$.
On the other hand, using the ilmenite lattice parameters at $x=66\%$ instead
of hematite (volume increase of 8.7\%) leads to an energy gain of
0.22~eV/p.f.u. for T$_{7,8,11,12}$ (red filled up-triangles in Fig.~\ref{fig:Ex}).

An interesting trend is observed in the shortest cation-oxygen bond lengths
(cf. Table~\ref{tab:enmm}), which tend to relax towards the values in the respective end member.
While $d_{\rm Fe^{3+}-O}$ (not shown) remains close to the value in bulk hematite (1.96\AA), the
bond lengths of the Ti-impurity and the neighboring \fetw\ relax towards the values in bulk ilmenite
(1.92 and 2.07\AA, respectively).

In Fe doped ilmenite the trend towards layered arrangements is retained for 66\%, \eg\ T$_{7,8,11,12}$
is favored by 0.21~eV compared to T$_{3,8,11,12}$, but ordered and disordered phases are
nearly degenerate at 83\%. Fe substituting for Ti in the
ilmenite lattice  is \feth. Additionally, one Fe in the neighboring layer turns \feth\
to compensate the charge, forming a \fetw, \feth contact layer. The substituted
Fe shows a very strong tendency to couple antiparallel to the neighboring Fe-layers.

In summary, we present a comprehensive GGA+U-study of the cation, charge and
magnetic order in the hematite-ilmenite system showing a strong preference
towards formation of layered configurations as opposed
to solid solutions. At the interface between hematite and ilmenite blocks we find evidence
for the \lmh~\cite{nature04} with a disproportionated \fetw, \feth contact layer to
accommodate the polar discontinuity. These uncompensated moments lead to ferrimagnetic behavior of the system.
The $d_{z^2}$ orbital in one spin-channel (all $d$ orbitals being occupied in the other)
at the \fetw-sites  in the contact layer  crosses the
Fermi level leading to halfmetallic behavior in most of the studied compositions.
 The ferrimagnetism
emerging at the interface of two antiferromagnetic oxides like hematite and ilmenite is an
impressive example of the novel functionality that can arise as a consequence of a polar discontinuity and the richer possibilities to compensate it that complex oxides offer. Recently,
an exchange bias of more than 1 Tesla was reported in this system~\cite{McEnroe07}. Further phenomena
such as oscillatory exchange coupling and spin-polarized transport remain to be explored in
controlled epitaxial \hem-\ilm-multilayers on the route to
possible device applications.

We acknowledge gratefully discussions with M. Winklhofer, W. Moritz and
R. Harrison as well as funding by the DFG (Pe883/4-1) and ESF within the
European Mineral Science Initiative (EuroMinSci). Simulations are performed
on the high performance supercomputer at the Leibniz Rechenzentrum.


\begin{thebibliography}{10}
\bibitem{U=6} The same halfmetallic behavior is obtained also for a lower $U$ value of 6~eV.
\end{thebibliography}
\end{document}